\documentclass[conference]{IEEEtran}
\usepackage{amssymb}
\usepackage{graphicx}
\graphicspath {{Figures/}}
\usepackage[table]{xcolor}
\usepackage{array}
\usepackage{listings}
\usepackage{fancyvrb}

\usepackage{comment}
\usepackage{url}
\usepackage{hyperref}
\usepackage{soul}

\ifCLASSINFOpdf
\else
\fi
\begin{document}
%
\title{Towards Object Life Cycle-Based\\ Variant Generation of Business Process Models}

\author{\IEEEauthorblockN{Ahmed Tealeb}
\IEEEauthorblockA{Information Systems Department,\\ Faculty of Computers and Information,\\
Cairo University, Egypt\\
Email: ahtealeb@gmail.com}}


%


\maketitle

\begin{abstract}
Variability management of process models is a major challenge for Process-Aware Information Systems. Process model variants can be attributed to any of the following reasons: new technologies, governmental rules, organizational context or adoption of new standards. Current approaches to manage variants of process models address issues such as reducing the huge effort of modeling from scratch, preventing redundancy, and controlling inconsistency in process models. Although the effort to manage process model variants has been exerted, there are still limitations. Furthermore, existing approaches do not focus on variants that come from change in organizational perspective of process models. Organizational-driven variant management is an important area that still needs more study that we focus on in this paper. Object Life Cycle (OLC) is an important aspect that may change from an organization to another. This paper introduces an approach inspired by real life scenario to generate consistent process model variants that come from adaptations in the OLC.
\end{abstract}

%
\IEEEpeerreviewmaketitle

\section{Introduction} \label{Introduction}
In recent years, the increasing adoption of Process-Aware Information Systems (PAISs) has resulted in large process model repositories [1]. One of the ongoing research challenges in the PAISs area is variability management. Each process variant constitutes an adjustment of a reference or basic process model to specific requirements. Efficient management for process model variants is a critical issue for organizations with the aim of helping them reduce the huge effort of modeling from scratch, prevent redundancy, and tackle inconsistency in process models. 
Despite the effort done in current approaches e.g., in Provop [2], C-EPCs [3], and PPM [4] to manage process model variants, there are still un-addressed issues. Current approaches focus on dealing with variants coming from change in control and behavioral perspectives of process models. However, variants originating from organizational and informational perspectives still need to be studied. 

\par The organizational perspective is one of the different views integrated in the process model. It identifies the hierarchy of the organization within which the process will be executed. Russell et al. \cite{6} introduced a set of Workflow Resource patterns (WRP) to capture the requirements for resource management such as representation and utilization in workflow environment \cite{6}. Awad et al. \cite{7} proposed an extension metamodel for BPMN to enable representation of resource assignment constraints using Object Constraints Language (OCL) \cite {8} to WRP \cite{6}. We, in a previous work, discussed organizational structures, and resource assignment matrix as aspects of the organizational perspective in [\cite{9}, \cite{10}]. 

\par Another important aspect of the organizational perspective is the Object Life Cycle (OLC). OLC enables organizations to understand the complete behavior of business objects. OLC is usually modeled using state machine \cite{14}. The aim of this paper is to propose a context-based approach for generating consistent process model variants focusing on the different variations of OLC. We enrich the OLC for a given base model by a set of exceptional cases extracted from UML Sequence Diagram. Then, we enable the user to select the suitable case. Finally, we generate the variant of the base model in hand.

\par The rest of this paper is organized as follows: Section \ref{Background} introduces the basic concepts related to our approach and discusses a motivating scenario. Section \ref{OLC-Based Variant Generation} presents our OLC-based algorithms for generating consistent process model variants. Section \ref{Related Work} discusses related work. Finally, Section \ref{Conclusions and Future Work} concludes our approach and outlines directions for future research.


 

\section{Background} \label{Background}
In this section, we introduce in brief the background that is related to our approach. We present the work of K\"{u}ster et al. \cite{15} for generating compliant business process model based on OLC and exception handling in both Business Process Model and Notation (BPMN) \cite{11} and UML Sequence Diagram (UMLSD) \cite{12} in the sections \ref{OLC-Based Process Model} and \ref{Exception Handling} respectively. Then, we present our motivating scenario in section \ref{Motivating Scenario}.

\subsection{OLC-Based Process Model} \label{OLC-Based Process Model}
\begin{figure}[!t]
    \centering
    \includegraphics[width=3.5in]{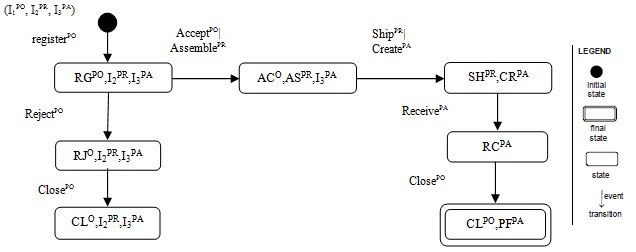}
    \caption{Composition of the Order, Product, and Payment OLCs}
    \label{Composition of OLCs}
\end{figure}

K\"{u}ster et al. \cite{15} introduced an approach for generating a compliant business process model from a set of  OLCs. Compliance using OLCs verifies the consistency of organization's processes and correct the execution of processes that spans several organizations. They formally defined the the compliance of a business process with an OLC using both OLC conformance and OLC coverage \cite{15}. OLC conformance happens when process model induced only the object state transitions which are defined in OLC. OLC coverage means that process model must cover the transitions and states in OLC. OLC Composition consists of an integrated view for a set of OLCs synchronized together. Fig.\ref{Composition of OLCs} represents a composition of Order, Product and payment OLCs. The OLC initiated by ($I_1^{PO}$, $I_2^{PR}$, $I_3^{PA}$). Firstly, the transition $register^{PO}$ registers the order with state \textquotedblleft RG\textquotedblright. Then, the order either accepted or rejected. If rejected, the transition $Reject^{PO}$ rejects the order with state \textquotedblleft RJ\textquotedblright, the order is closed as a result of order's rejection. If accepted, the transition $Accept^{PO}$$|$$Assemble^{PR}$ will accept the order and assemble the object product with the state \textquotedblleft $AC^{PO}$, $AS^{PR}$, $I_3^{PA}$\textquotedblright. Then, the transition $Ship^{PR}$$|$$Creater^{PA}$ ships the products with the state \textquotedblleft $SH^{PR}$, $CR^{PA}$\textquotedblright. Then, the transition $Receive^{PA}$ changes the state to \textquotedblleft $RC^{PA}$\textquotedblright. Then, the order is closed as a result of receiving the payment.

The aim of K\"{u}ster et al. \cite{15} work is introducing a novel approach to generate a compliant process model using OLC. However, our approach addresses the issue of generating consistent process model variants from exceptional cases based on context that are not covered in OLC.

\subsection{Exception Handling} \label{Exception Handling}
Each business process may be exposed by an exception that deviate a process from its normal execution flow. In general, Exceptions are classified into two types: \textit {business related} (e.g. customer cancels  order late), and \textit{IT related} (e.g.system errors or crashes) \cite{16}. In BPM, we classify exceptions into three types: \textit {external} (i.e. something goes wrong outside the process), \textit{internal} (i.e. something goes wrong inside the process), and \textit{timeout} (i.e. activity takes a longtime) \cite{16}.

\par UML Sequence Diagram is the most common type of interaction diagrams. It is used to show the interactions between objects arranged in time sequence \cite{12}. The \textit {break combined fragment} are used to model exception handling \cite{13}.

\par We make use of these exceptional cases as adaptations in our approach to generate variant process models for the base process model in Section \ref{Motivating Scenario}.

\subsection{Motivating Scenario} \label{Motivating Scenario}
In this section, we introduce a real life scenario that motivated the development of our approach. We introduce the base model for \textquotedblleft Oder Fulfillment\textquotedblright \hspace{0cm} business process based on OLC in Fig. \ref{Composition of OLCs}.

\begin{figure}[ht!]
    \centering
    \includegraphics [width=0.9\linewidth]{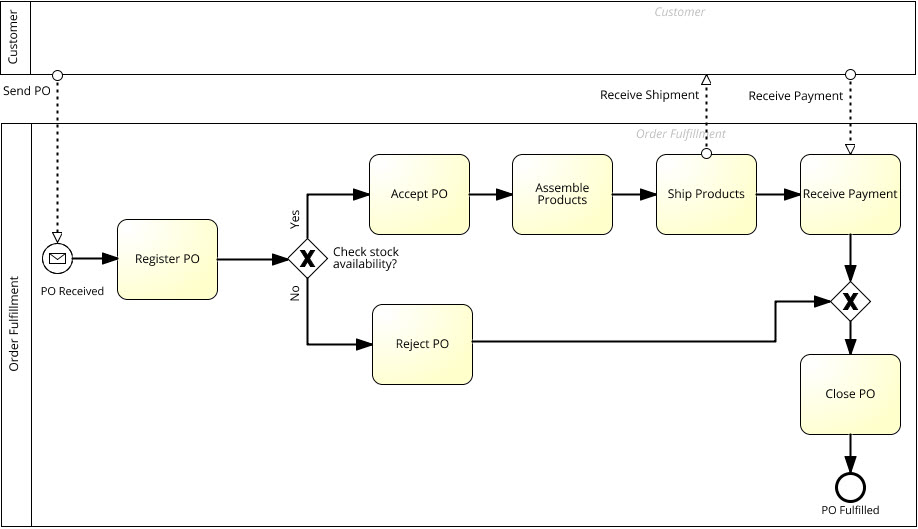}
    \caption{Order Fulfillment - Base Process Model}
    \label{Order Fulfillment - Base Process Model}
\end{figure}

Oder Fulfillment is one of the most frequently executed business processes in organizations. The process starts when a customer sends a Purchase Order (PO) to an organization. Then, the organization registers the PO after receiving it. Then, it checks the stock for the availability. If the products are not available, the organization rejects the PO. Otherwise, the organization accepts the PO, assembles the products, ship products and finally receives payment from the customer. Fig. \ref{Order Fulfillment - Base Process Model} represents a \textit{base model} for the \textquotedblleft Oder Fulfillment\textquotedblright \hspace{0cm} process.

The Order Fulfillment process may vary from one organization to another. We discuss in detail how these variants may be generated in section \ref{OLC-Based Variant Generation}.

\section{OLC-Based Variant Generation} \label{OLC-Based Variant Generation}
In this section, we introduce a solution for generating consistent process model variants based on OLC and Exception Handling mentioned in Section \ref{Background}. We present two OLC-based algorithms as follows: \textquotedblleft \textit{OLC Adaptations}\textquotedblright \hspace{0cm} and \textquotedblleft \textit {OLC-Based Process Variant Generator}\textquotedblright \hspace{0cm} in the sections \ref{OLC Adaptations}, and \ref{OLC-Based Process Variant Generator} respectively. 

\subsection{OLC Adaptations} \label{OLC Adaptations}
\begin{figure}[ht!]
    \centering
    \includegraphics [width=0.75\linewidth]{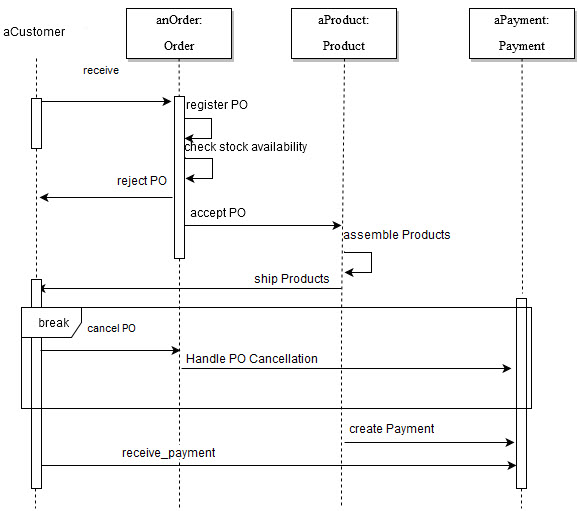}
    \caption{Order Fulfillment - UML Sequence Diagram}
    \label{Order Fulfillment - UML Sequence Diagram}
\end{figure}

The \textquotedblleft OLC Adaptations\textquotedblright \hspace{0cm} algorithm is responsible for retrieving the exceptional cases which are not covered by OLC for a given base process model as in Fig. \ref{Order Fulfillment - Base Process Model}. Firstly, OLC Adaptations starts with reading the different \textit{break combined fragments} from UML Sequence Diagram as in Fig. \ref{Order Fulfillment - UML Sequence Diagram}. Secondly, we enable the user to select the exceptional case suits the context.

\noindent \textbf{Algorithm 3.1 OLC Adaptations.}\\
\textbf{Inputs:} \textit{OLC} is the object life cycle of the base model, \textit{UMLSD} is the UML Sequence Diagram contains break combined fragment(s).\\
\textbf{Outputs:} \textit{AOLC} is the adapted OLC with a selected break combined fragment.\\
\textbf{Variables:} \textit{BCF[]} is an array of break combined fragments, \textit{BCFInitiator} is the initiator type for BCF such as \textit{Internal, External, or Timeout} Exception, \textit{BCFTrans} are set of transitions , \textit{PT} represents previous transition for BCF, \textit{NT} represents next transition for BCF, \textit{P} is a specific position in AOLC.

\begin{Verbatim}[numbers=left]
AOLC = OLC;
For each BCF in UMLSD
  BCF[] = AddBcf(BCF)
  For each BCF in BCF[]
  BCFInitiator = BCF.getInitiatorType()
  BCFTrans = BCF.getTransition()
  PT = UMLSD.getPreviousTransition(BCF)
  NT = UMLSD.getNextTransition(BCF)
  P = AOLC.getPosition(PT,NT)
  AOLC = InsertBcf(P, BCFTrans, BCFInitiator)
  End For
End For
return AOLC
\end{Verbatim}
Line 1 initiates AOLC by the OLC of the base process model. Line 2-3 reads all BCFs in the UMLSD and insert them into array of \textit{break combined fragments}. Lines 4-11 perform a number of actions for each BCF. Firstly, we get the type of initiator for BCF such as \textit{Internal, External, or Timeout} Exception in Line 5. Secondly, we get the transitions in BCF in Line 6. Thirdly, we get both previous \textquotedblleft PT\textquotedblright \hspace{0cm} and next \textquotedblleft NT\textquotedblright \hspace{0cm} transitions for the BCF using the UMLSD in Lines 7-8 respectively. Fourthly, we get the position inside AOLC using \textquotedblleft PT\textquotedblright \hspace{0cm} and \textquotedblleft NT\textquotedblright \hspace{0cm} transitions for the BCF in Line 9. Fifthly, we update AOLC by BCFTrans and BCFInitiator  in Line 10. Finally, we return the AOLC in Line 13.

\subsection{OLC-Based Process Variant Generator} \label{OLC-Based Process Variant Generator}
The \textquotedblleft OLC-Based Process Variant Generator\textquotedblright \hspace{0cm} algorithm manages the issue of generating consistent process model variants caused by adaptations in OLC generated by \textquotedblleft OLC Adaptations\textquotedblright \hspace{0cm} algorithm in section \ref{OLC Adaptations}.\\

\noindent \textbf{Algorithm 3.2 OLC-Based Process Variant Generator.}\\
\textbf{Inputs:} \textit{BM} is the base model of a process; \textit{AOLC} is the adapted OLC with a selected break combined fragment.\\
\textbf{Outputs:} \textit{VPM} variant process model.\\
\textbf{Variables:} \textit{OPR} represents the operation based on BCFInitiator of AOLC, \textit{EEP} is external exception pattern, \textit{IEP} is internal exception pattern, \textit{TEP} is timeout exception pattern, \textit{VPM} initially is the base model.

\begin{figure}[ht!]
    \centering
    \includegraphics [width=0.85\linewidth]{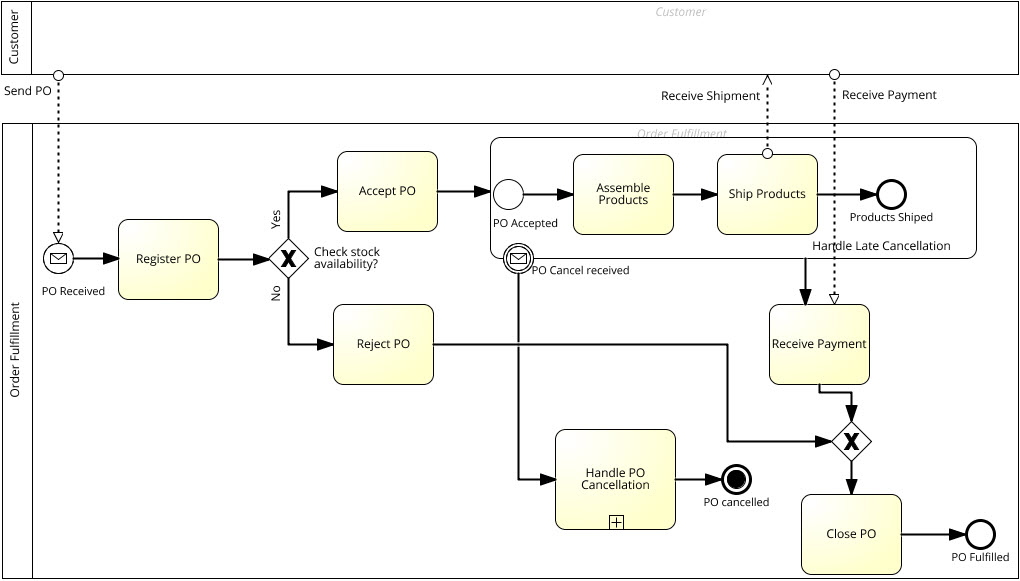}
    \caption{Order Fulfillment - Late Cancellation Variant Process Model}
    \label{Order Fulfillment - Late Cancellation Variant Process Model}
\end{figure}

\begin{Verbatim}[numbers=left]
VPM = BM;
OPR = AOLC.getBcfInitiator()
If OPR = ‘Insert External Exception’ Then
  VPM.insertPattern(AOLC,EEP)
Else If OPR = ‘Insert Internal Exception’ Then
  VPM.insertPattern(AOLC,IEP)
Else If OPR = ‘Insert Timeout Exception’ Then 
  VPM.insertPattern(AOLC,TEP)
End If
return VPM
\end{Verbatim}
Line 1 defines and initializes VPM to BM. Line 2 retrieves the operation of AOLC. Then, the algorithm checks the type of OPR. Lines 3-4 if correct, algorithm performs an insert external exception pattern. Lines 5-6 if correct, algorithm performs an insert internal exception pattern. Lines 7-8 if correct, algorithm performs an insert timeout exception pattern. Line 10 returns the generated VPM. Fig. \ref{Order Fulfillment - Late Cancellation Variant Process Model} shows the generated variant of the base model in Fig. \ref{Order Fulfillment - Base Process Model}.

\par So, we can conclude that the common source for \textit{the variant} introduced for the process of \textquotedblleft Order Fulfillment\textquotedblright \hspace{0cm} before is the changes in OLC.

\section{Related Work} \label{Related Work}
Several approaches have been developed in recent years to manage the different variants of process models, Such as PROcess Variants by OPtions (Provop), Configurable Event-driven Process Chains (C-EPCs), and Partial Process Models (PPM). In this section, we state the pros and cons for each approach.
\par Provop is an approach for managing a set of related process variants throughout the entire Business Process Life Cycle (BPLC) \cite{3}. In Provop, a specific variant is derived by adjusting the basic process model using a set of well-defined change operations \cite{3}. Change Operations represent the difference between basic model and variant such as INSERT, DELETE, and MOVE process fragments, and MODIFY process elements attributes. Furthermore, Provop supports the context-aware process configuration either statically or dynamically \cite{17}.  The Provop lifecycle \cite{18} consists of three major phases: the modeling phase, the configuration phase and the execution phase. Provop has been extended with a procedure to guarantee the correctness and soundness for a \textit{family of configurable process variants} \cite{19}. An extension has been developed for ARIS Business Architect to cope with variability in process models based on Provop \cite{2}. Provop uses a bottom-up technique from process variants to the basic process model. Each variant is maintained through the base model only. So, the changes in any variant may not be consistent with other variants of the same process.\\
The concept of configurable process model has been defined by \cite{4}. It merges variants of process models into a single configurable model. Configurable process models are integrated representations for variants of a process model in a specific domain. A framework to manage the configuration of business process models consists of three parts: \textit{a conceptual foundation for process model configuration, a questionnaire-based approach for validating modeling, and a meta-model for holistic process configuration} \cite{20}. C-EPCs are configurable version of EPCs, which provides a means to capture variability in EPC process models. C-EPCs identify a set of variation points which are called \textit{configurable nodes} in the model and constraints, which are called \textit{configuration requirements} to restrict the different combinations of allowed variants in order to be assigned for variants called \textit{alternatives} \cite{20}. La Rosa et al. \cite{21} proposed C-iEPC, that extends C-EPC notation with the notions of roles and objects associated to functions. C-iEPC supports variants from organizational perspective. C-EPCs uses a top-down technique from holistic or reference process model to process variants. Specifying all variants in a holistic reference model for a particular process is difficult to maintain.\\
\indent PPM is a query-based approach that depends on defining process model views to maintain consistency among process variants \cite{5}. These views are defined using a visual query language for business process models called BPMN-Q \cite{22}. Based on BPMN-Q, a framework for querying and reusing business process models has been developed by \cite{23}. PPM is using inheritance mechanisms from software engineering to make best use of the reusability as a concept of Object-Oriented Modeling of object orientation \cite{5}. PPM approach provides support for consistency of process model variants, and allows handling issues for multiple inheritance levels. PPM uses both top-down and bottom-up techniques in handling process variants. Context issues related to variants of business process are not covered in the PPM approach.\\
\indent Despite the significant effort that has gone into the current approaches to manage process models variants, the organizational perspective has many aspects still need to be studied. So, Our approach focus on OLC as one important aspect of organizational perspective to manage generating the process model variants consistently.

\section{Conclusions and Future Work} \label{Conclusions and Future Work}
This paper introduces an approach to manage the generation of consistent process model variants that come from adaptations in OLC. The approach helps practitioners, such as process owners and/or designers in generating consistent variants of their process models depending on the case in hand. The most significant finding behind the approach is the importance of the organizational perspective's aspects such as OLC. In this paper, we presented two context-based algorithm to derive variants of process models based on OLC. We applied the approach to real life process models to further illustrate our ideas.
\par In future work, we seek to apply the approach for more real world cases in different domains. Furthermore, we look for other aspects from the organizational perspective to complete our approach. Finally, a proof-of-concept prototype that validates the concept behind approach will be implemented.






%

\end{document}